\documentclass[useAMS,usenatbib]{mn2e}
\usepackage{tabularx}
\usepackage{xspace}
\usepackage{graphicx}
\newcommand{\ignore}[1]{}

\providecommand{\ao}{}

\renewcommand{\ao}{adaptive optics (AO)\renewcommand{\ao}{AO\xspace}\renewcommand{\Ao}{AO\xspace}\xspace}
\newcommand{\Ao}{Adaptive optics (AO)\renewcommand{\ao}{AO\xspace}\renewcommand{\Ao}{AO\xspace}\xspace}
\newcommand{\wfs}{wavefront sensor (WFS)\renewcommand{\wfs}{WFS\xspace}\renewcommand{\wfss}{WFSs\xspace}\xspace}
\newcommand{\wfss}{wavefront sensors (WFSs)\renewcommand{\wfs}{WFS\xspace}\renewcommand{\wfss}{WFSs\xspace}\xspace}

\newcommand{\hpc}{high performance computing (HPC)\renewcommand{\hpc}{HPC\xspace}\xspace}
\newcommand{\iac}{Instituto de Astrofisica de Canarias (IAC)\renewcommand{\iac}{IAC\xspace}\xspace}
\newcommand{\gmt}{Giant Magellan Telescope (GMT)\renewcommand{\gmt}{GMT\xspace}\xspace}
\newcommand{\lsst}{Large Synoptic Sky Telescope (LSST)\renewcommand{\lsst}{LSST\xspace}\xspace}
\newcommand{\ska}{Square Kilometer Array (SKA)\renewcommand{\ska}{SKA\xspace}\xspace}

\newcommand{\shwfs}{Shack-Hartmann \wfs (SHWFS)\renewcommand{\shwfs}{SHWFS\xspace}\xspace}
\newcommand{\dm}{deformable mirror (DM)\renewcommand{\dm}{DM\xspace}\renewcommand{\dms}{DMs\xspace}\renewcommand{\Dms}{DMs\xspace}\renewcommand{\Dm}{DM\xspace}\xspace}
\newcommand{\dms}{deformable mirrors (DMs)\renewcommand{\dm}{DM\xspace}\renewcommand{\dms}{DMs\xspace}\renewcommand{\Dms}{DMs\xspace}\renewcommand{\Dm}{DM\xspace}\xspace}
\newcommand{\Dms}{Deformable mirrors (DMs)\renewcommand{\dm}{DM\xspace}\renewcommand{\dms}{DMs\xspace}\renewcommand{\Dms}{DMs\xspace}\renewcommand{\Dm}{DM\xspace}\xspace}
\newcommand{\Dm}{Deformable mirror (DM)\renewcommand{\dm}{DM\xspace}\renewcommand{\dms}{DMs\xspace}\renewcommand{\Dms}{DMs\xspace}\renewcommand{\Dm}{DM\xspace}\xspace}
\newcommand{\fov}{field of view (FOV)\renewcommand{\fov}{FOV\xspace}\xspace}
\newcommand{\pol}{pseudo open loop (POL)\renewcommand{\pol}{POL\xspace}\xspace}
\newcommand{\lqg}{linear-quadratic-gaussian (LQG)\renewcommand{\lqg}{LQG\xspace}\xspace}
\newcommand{\shs}{Shack-Hartmann sensor (SHS)\renewcommand{\shs}{SHS\xspace}\renewcommand{\shss}{SHSs\xspace}\xspace}
\newcommand{\shss}{Shack-Hartmann sensors (SHSs)\renewcommand{\shs}{SHS\xspace}\renewcommand{\shss}{SHSs\xspace}\xspace}
\newcommand{\lgs}{laser guide star (LGS)\renewcommand{\lgs}{LGS\xspace}\renewcommand{\Lgs}{LGS\xspace}\renewcommand{\lgss}{LGSs\xspace}\xspace}
\newcommand{\lgss}{laser guide stars (LGSs)\renewcommand{\lgs}{LGS\xspace}\renewcommand{\Lgs}{LGS\xspace}\renewcommand{\lgss}{LGSs\xspace}\xspace}
\newcommand{\Lgs}{Laser guide star (LGS)\renewcommand{\lgs}{LGS\xspace}\renewcommand{\Lgs}{LGS\xspace}\renewcommand{\lgss}{LGSs\xspace}\xspace}

\newcommand{\Ngs}{Natural guide star (NGS)\renewcommand{\ngs}{NGS\xspace}\renewcommand{\Ngs}{NGS\xspace}\renewcommand{\ngss}{NGSs\xspace}\xspace}
\newcommand{\ngs}{natural guide star (NGS)\renewcommand{\ngs}{NGS\xspace}\renewcommand{\Ngs}{NGS\xspace}\renewcommand{\ngss}{NGSs\xspace}\xspace}
\newcommand{\ngss}{natural guide stars (NGSs)\renewcommand{\ngs}{NGS\xspace}\renewcommand{\Ngs}{NGS\xspace}\renewcommand{\ngss}{NGSs\xspace}\xspace}
\newcommand{\mems}{Micro-Electro-Mechanical Systems (MEMS)\renewcommand{\mems}{MEMS\xspace}\xspace}
\newcommand{\snr}{signal to noise ratio (SNR)\renewcommand{\snr}{SNR\xspace}\xspace}
\newcommand{\Moao}{Multi-object \ao (MOAO)\renewcommand{\moao}{MOAO\xspace}\renewcommand{\Moao}{MOAO\xspace}\xspace}
\newcommand{\moao}{multi-object \ao (MOAO)\renewcommand{\moao}{MOAO\xspace}\renewcommand{\Moao}{MOAO\xspace}\xspace}
\newcommand{\mcao}{multi-conjugate adaptive optics (MCAO)\renewcommand{\mcao}{MCAO\xspace}\xspace}
\newcommand{\ltao}{laser tomographic \ao (LTAO)\renewcommand{\ltao}{LTAO\xspace}\xspace}
\newcommand{\cpu}{central processing unit (CPU)\renewcommand{\cpu}{CPU\xspace}\renewcommand{\cpus}{CPUs\xspace}\xspace}
\newcommand{\cpus}{central processing units (CPUs)\renewcommand{\cpu}{CPU\xspace}\renewcommand{\cpus}{CPUs\xspace}\xspace}
\newcommand{\psf}{point spread function (PSF)\renewcommand{\psf}{PSF\xspace}\renewcommand{\psfs}{PSFs\xspace}\renewcommand{\Psf}{PSF\xspace}\xspace}
\newcommand{\psfs}{point spread functions (PSFs)\renewcommand{\psf}{PSF\xspace}\renewcommand{\psfs}{PSFs\xspace}\renewcommand{\Psf}{PSF\xspace}\xspace}
\newcommand{\Psf}{Point spread function (PSF)\renewcommand{\psf}{PSF\xspace}\renewcommand{\psfs}{PSFs\xspace}\renewcommand{\Psf}{PSF\xspace}\xspace}
\newcommand{\fpga}{field programmable gate array (FPGA)\renewcommand{\fpga}{FPGA\xspace}\renewcommand{\fpgas}{FPGAs\xspace}\xspace}
\newcommand{\fpgas}{field programmable gate arrays (FPGAs)\renewcommand{\fpga}{FPGA\xspace}\renewcommand{\fpgas}{FPGAs\xspace}\xspace}
\newcommand{\sor}{successive over-relaxation (SOR)\renewcommand{\sor}{SOR\xspace}\xspace}
\newcommand{\fdpcg}{Fourier domain pre-conditioned gradient (FDPCG)\renewcommand{\fdpcg}{FDPCG\xspace}\xspace}
\newcommand{\map}{maximum a-posteriori (MAP)\renewcommand{\map}{MAP\xspace}\xspace}

\newcommand{\elt}{Extremely Large Telescope (ELT)\renewcommand{\elt}{ELT\xspace}\renewcommand{\elts}{ELTs\xspace}\renewcommand{\eelt}{European ELT (E-ELT)\renewcommand{\eelt}{E-ELT\xspace}\xspace}\xspace}

\newcommand{\elts}{Extremely Large Telescopes (ELTs)\renewcommand{\elt}{ELT\xspace}\renewcommand{\elts}{ELTs\xspace}\renewcommand{\eelt}{European ELT (E-ELT)\renewcommand{\eelt}{E-ELT\xspace}\xspace}\xspace}

\newcommand{\eelt}{European Extremely Large Telescope (E-ELT)\renewcommand{\eelt}{E-ELT\xspace}\renewcommand{\elt}{ELT\xspace}\renewcommand{\elts}{ELTs\xspace}\xspace}

\newcommand{\dugall}{Durham University generalised adaptive optics laser laboratory (DUGALL)\renewcommand{\dugall}{DUGALL\xspace}\xspace}
\newcommand{\fwhm}{full-width at half-maximum (FWHM)\renewcommand{\fwhm}{FWHM\xspace}\xspace}
\newcommand{\wht}{William Herschel Telescope (WHT)\renewcommand{\wht}{WHT\xspace}\xspace}
\newcommand{\emccd}{electron multiplying CCD (EMCCD)\renewcommand{\emccd}{EMCCD\xspace}\renewcommand{\emccds}{EMCCDs\xspace}\xspace}
\newcommand{\emccds}{electron multiplying CCDs (EMCCDs)\renewcommand{\emccd}{EMCCD\xspace}\renewcommand{\emccds}{EMCCDs\xspace}\xspace}
\newcommand{\dasp}{Durham \ao simulation platform (DASP)\renewcommand{\dasp}{DASP\xspace}\renewcommand{\thedasp}{DASP\xspace}\renewcommand{\Thedasp}{DASP\xspace}\renewcommand{\daspcite}{DASP\xspace}\renewcommand{\daspcite}{DASP\xspace}\renewcommand{\thedaspcite}{DASP\xspace}\xspace}
\newcommand{\daspcite}{Durham \ao simulation platform \citep[DASP,]{basden5}\renewcommand{\dasp}{DASP\xspace}\renewcommand{\thedasp}{DASP\xspace}\renewcommand{\Thedasp}{DASP\xspace}\renewcommand{\daspcite}{DASP\xspace}\renewcommand{\thedaspcite}{DASP\xspace}\xspace}
\newcommand{\thedaspcite}{the Durham \ao simulation platform \citep[DASP,]{basden5}\renewcommand{\dasp}{DASP\xspace}\renewcommand{\thedasp}{DASP\xspace}\renewcommand{\Thedasp}{DASP\xspace}\renewcommand{\daspcite}{DASP\xspace}\renewcommand{\thedaspcite}{DASP\xspace}\xspace}

\newcommand{\thedasp}{the Durham \ao simulation platform (DASP)\renewcommand{\dasp}{DASP\xspace}\renewcommand{\thedasp}{DASP\xspace}\renewcommand{\Thedasp}{DASP\xspace}\renewcommand{\daspcite}{DASP\xspace}\renewcommand{\thedaspcite}{DASP\xspace}\xspace}
\newcommand{\Thedasp}{The Durham \ao simulation platform (DASP)\renewcommand{\dasp}{DASP\xspace}\renewcommand{\thedasp}{DASP\xspace}\renewcommand{\Thedasp}{DASP\xspace}\renewcommand{\daspcite}{DASP\xspace}\renewcommand{\thedaspcite}{DASP\xspace}\xspace}
\newcommand{\mpi}{Message Passing Interface (MPI)\renewcommand{\mpi}{MPI\xspace}\xspace}
\newcommand{\smp}{symmetric multi-processing (SMP)\renewcommand{\smp}{SMP\xspace}\xspace}
\newcommand{\svd}{singular value decomposition (SVD)\renewcommand{\svd}{SVD\xspace}\xspace}
\newcommand{\gpu}{graphics processing unit (GPU)\renewcommand{\gpu}{GPU\xspace}\renewcommand{\gpus}{GPUs\xspace}\xspace}
\newcommand{\gpus}{graphics processing units (GPUs)\renewcommand{\gpu}{GPU\xspace}\renewcommand{\gpus}{GPUs\xspace}\xspace}
\newcommand{\fft}{fast Fourier transform (FFT)\renewcommand{\fft}{FFT\xspace}\xspace}
\newcommand{\ifu}{integral field unit (IFU)\renewcommand{\ifu}{IFU\xspace}\xspace}
\newcommand{\darc}{the Durham \ao real-time controller (DARC)\renewcommand{\darc}{DARC\xspace}\renewcommand{\Darc}{DARC\xspace}\xspace}
\newcommand{\Darc}{The Durham \ao real-time controller (DARC)\renewcommand{\darc}{DARC\xspace}\renewcommand{\Darc}{DARC\xspace}\xspace}
\newcommand{\darccite}{the Durham \ao real-time controller \citep[DARC,]{basden9}\renewcommand{\darc}{DARC\xspace}\renewcommand{\Darc}{DARC\xspace}\renewcommand{\darccite}{DARC\xspace}\xspace}
\newcommand{\cots}{commercial off-the-shelf (COTS)\renewcommand{\cots}{COTS\xspace}\xspace}
\newcommand{\rtcp}{real-time control pipeline (RTCP)\renewcommand{\rtcp}{RTCP\xspace}\xspace}
\newcommand{\rms}{root-mean-square (RMS)\renewcommand{\rms}{RMS\xspace}\xspace}
\newcommand{\sFPDP}{serial Front Panel Data Port (sFPDP)\renewcommand{\sFPDP}{sFPDP\xspace}\xspace}
\newcommand{\wpu}{wavefront processing unit (WPU)\renewcommand{\wpu}{WPU\xspace}\xspace}

\newcommand{\rtcs}{real-time control system (RTCS)\renewcommand{\rtcs}{RTCS\xspace}\renewcommand{\rtcss}{RTCSs\xspace}\xspace}
\newcommand{\rtcss}{real-time control systems (RTCSs)\renewcommand{\rtcs}{RTCS\xspace}\renewcommand{\rtcss}{RTCSs\xspace}\xspace}
\newcommand{\eso}{European Southern Observatory (ESO)\renewcommand{\eso}{ESO\xspace}\renewcommand{\theeso}{ESO\xspace}\xspace}
\newcommand{\theeso}{\renewcommand{\theeso}{ESO\xspace}the \eso}
\newcommand{\scao}{single conjugate \ao (SCAO)\renewcommand{\scao}{SCAO\xspace}\renewcommand{\Scao}{SCAO\xspace}\xspace}
\newcommand{\Scao}{Single conjugate \ao (SCAO)\renewcommand{\scao}{SCAO\xspace}\renewcommand{\Scao}{SCAO\xspace}\xspace}
\newcommand{\glao}{ground layer \ao (GLAO)\renewcommand{\glao}{GLAO\xspace}\xspace}
\newcommand{\eagle}{ELT Adaptive optics for GaLaxy Evolution (EAGLE)\renewcommand{\eagle}{EAGLE\xspace}\xspace}
\newcommand{\maory}{multi-conjugate \ao relay for the \eelt (MAORY)\renewcommand{\maory}{MAORY\xspace}\xspace}
\newcommand{\muse}{Multi Unit Spectroscopic Explorer (MUSE)\renewcommand{\muse}{MUSE\xspace}\xspace}
\newcommand{\vlt}{Very Large Telescope (VLT)\renewcommand{\vlt}{VLT\xspace}\xspace}

\newcommand{\eapd}{electron avalanche photodiode\renewcommand{\eapd}{eAPD\xspace}\xspace}
\newcommand{\tmt}{Thirty Metre Telescope (TMT)\renewcommand{\tmt}{TMT\xspace}\xspace}
\newcommand{\lbt}{Large Binocular Telescope (LBT)\renewcommand{\lbt}{LBT\xspace}\xspace}
\newcommand{\xao}{eXtreme \ao (XAO)\renewcommand{\xao}{XAO\xspace}\renewcommand{\Xao}{XAO\xspace}\xspace}
\newcommand{\Xao}{EXtreme \ao (XAO)\renewcommand{\xao}{XAO\xspace}\renewcommand{\Xao}{XAO\xspace}\xspace}

\newcommand{\pcs}{Planetary Camera and Spectrograph (PCS)\renewcommand{\pcs}{PCS\xspace}\xspace}
\newcommand{\vla}{Very Large Array (VLA)\renewcommand{\vla}{VLA\xspace}\xspace}
\newcommand{\jwst}{{\em James Webb Space Telescope} \citep[JWST,][]{jwst}\renewcommand{\jwst}{{\em JWST}\xspace}\xspace}
\newcommand{\hst}{{\em Hubble Space Telescope (HST)}\renewcommand{\hst}{{\em HST}\xspace}\xspace}
\newcommand{\ifss}{integral-field spectrographs (IFSs)\renewcommand{\ifss}{IFSs\xspace}\renewcommand{\ifs}{IFS\xspace}\xspace}
\newcommand{\ifs}{integral-field spectrograph (IFS)\renewcommand{\ifss}{IFSs\xspace}\renewcommand{\ifs}{IFS\xspace}\xspace}
\newcommand{\ifus}{integral field units (IFUs)\renewcommand{\ifus}{IFUs\xspace}\xspace}
\newcommand{\mos}{multi-object spectrograph (MOS)\renewcommand{\mos}{MOS\xspace}\xspace}
\newcommand{\goodss}{Great Observatories Origins Deep Survey (GOODS)-S\renewcommand{\goodss}{GOODS-S\xspace}\xspace}
\newcommand{\goods}{Great Observatories Origins Deep Survey (GOODS)\renewcommand{\goods}{GOODS\xspace}\xspace}
\newcommand{\cmos}{complimentary metal-oxide semiconductor (CMOS)\renewcommand{\cmos}{CMOS\xspace}\xspace}
\newcommand{\scmos}{scientific CMOS (sCMOS)\renewcommand{\scmos}{sCMOS\xspace}\xspace}
\newcommand{\aof}{Adaptive Optics Facility (AOF)\renewcommand{\aof}{AOF\xspace}\xspace}
\newcommand{\dsp}{digital signal processor (DSP)\renewcommand{\dsp}{DSP\xspace}\renewcommand{\dsps}{DSPs\xspace}\xspace}
\newcommand{\dsps}{digital signal processors (DSPs)\renewcommand{\dsp}{DSP\xspace}\renewcommand{\dsps}{DSPs\xspace}\xspace}
\newcommand{\capi}{Coherent Accelerator Processor Interface (CAPI)\renewcommand{\capi}{CAPI\xspace}\xspace}
\newcommand{\qe}{quantum efficiency (QE)\renewcommand{\qe}{QE\xspace}\xspace}
\newcommand{\numa}{non-uniform memory access (NUMA)\renewcommand{\numa}{NUMA\xspace}\xspace}
\newcommand{\uav}{unmanned aerial vehicle (UAV)\renewcommand{\uav}{UAV\xspace}\renewcommand{\uavs}{UAVs\xspace}\renewcommand{\Uav}{UAV\xspace}\xspace}
\newcommand{\uavs}{unmanned aerial vehicles (UAVs)\renewcommand{\uav}{UAV\xspace}\renewcommand{\uavs}{UAVs\xspace}\renewcommand{\Uav}{UAV\xspace}\xspace}
\newcommand{\Uav}{Unmanned aerial vehicle (UAV)\renewcommand{\uav}{UAV\xspace}\renewcommand{\Uav}{UAV\xspace}\renewcommand{\uavs}{UAVs\xspace}\xspace}

\newcommand{\ncpa}{non-common path aberration (NCPA)\renewcommand{\ncpa}{NCPA\xspace}\renewcommand{\ncpas}{NCPAs\xspace}\xspace}
\newcommand{\ncpas}{non-common path aberrations (NCPA)\renewcommand{\ncpa}{NCPA\xspace}\renewcommand{\ncpas}{NCPAs\xspace}\xspace}
\newcommand{\sdk}{software developers kit (SDK)\renewcommand{\sdk}{SDK\xspace}\renewcommand{\sdks}{SDKs\xspace}\xspace}
\newcommand{\sdks}{software developers kits (SDKs)\renewcommand{\sdk}{SDK\xspace}\renewcommand{\sdks}{SDKs\xspace}\xspace}
\newcommand{\dac}{digital to analogue converter (DAC)\renewcommand{\dac}{DAC\xspace}\xspace}
\newcommand{\nda}{non-disclosure agreement (NDA)\renewcommand{\nda}{NDA\xspace}\xspace}
\newcommand{\polc}{pseudo-open-loop control (POLC)\renewcommand{\polc}{POLC\xspace}\xspace}
\newcommand{\udp}{User Datagram Protocol (UDP)\renewcommand{\udp}{UDP\xspace}\xspace}
\newcommand{\ags}{artificial guide star (AGS)\renewcommand{\ags}{AGS\xspace}\xspace}
\newcommand{\est}{European Solar Telescope (EST)\renewcommand{\est}{EST\xspace}\xspace}
\newcommand{\lot}{Large Optical Telescope (LOT)\renewcommand{\lot}{LOT\xspace}\xspace}
\newcommand{\gtc}{Gran Telescopio Canarias (GTC)\renewcommand{\gtc}{GTC\xspace}\xspace}
\newcommand{\cta}{Cherenkov Telescope Array (CTA)\renewcommand{\cta}{CTA\xspace}\xspace}
\newcommand{\rtk}{Real-time Kinematic (RTK)\renewcommand{\rtk}{RTK\xspace}\xspace}
\newcommand{\gnss}{global navigation satellite systems (GNSS)\renewcommand{\gnss}{GNSS\xspace}\xspace}
\newcommand{\race}{Rapid Automatic Cascode Exchange (RACE)\renewcommand{\race}{RACE\xspace}\xspace}
\newcommand{\tvm}{total variation minimisation (TVM)\renewcommand{\tvm}{TVM\xspace}\xspace}
\newcommand{\ccs}{Central Control System (CCS)\renewcommand{\ccs}{CCS\xspace}\xspace}

\usepackage{amssymb}
\usepackage{amsmath}
\title[Efficient POLC for ELTs]{Efficient implementation of
  pseudo open loop control for adaptive optics on Extremely Large Telescopes}

\author[A.\ G.\ Basden et al.]{A.\ G.\ Basden$^{1}$\thanks{E-mail:
    a.g.basden@durham.ac.uk (AGB)}, D.\ Jenkins$^{1}$, T.\ J.\ Morris$^{1}$,
  J.\ Osborn$^{1}$ and M.\ J.\ Townson$^{1}$.
\\
$^{1}$Centre for Advanced Instrumentation, Department of Physics, South Road, Durham, DH1 3LE,
UK
}
\begin{document}
\maketitle

\begin{abstract}
Closed-loop adaptive optics systems which use minimum mean square
error wavefront reconstruction require the computation of
pseudo open loop wavefront slopes.  These techniques incorporate a
knowledge of atmospheric statistics which must therefore be
represented within the wavefront slope measurements.  These
pseudo open loop slopes are computed from the sum of the measured
residual slopes and the reconstructed slopes that would be given if
the deformable mirror was flat, generally involving the multiplication
of an interaction matrix with actuator demands from the previous
time-step.  When using dense algebra, this multiplication is
computationally expensive for Extremely Large Telescopes, requiring a
large memory bandwidth.  Here we show that this requirement can be
significantly reduced, maintaining mathematical correctness and
significantly reducing system complexity.  This therefore reduces the
cost of these systems and increases robustness and reliability.
\end{abstract}


\section{Introduction}
\Ao \citep{adaptiveoptics,hardy} is now a mainstream technology, and
essential for the next generation of \elts, including the \eso \elt,
the \tmt and the \gmt, which will have light collecting areas
equivalent to primary mirror diameters of at least 20~m.  To optimise
\ao performance, particularly for wide \fov systems and at low signal
levels, it is necessary to use minimum variance wavefront
reconstruction techniques \citep{map}, also known as minimum mean
square error or maximum a posteriori reconstruction.  These methods
make use of atmospheric statistics to optimise the wavefront
reconstruction.  However, to do this, they must rely on having
open-loop slope measurements as input, i.e.\ those that represent the
statistics of the atmosphere, rather than the residual slope
measurements more commonly used by \scao systems.  This can be
achieved in two ways.  Either the wavefront sensors can be operated in
open-loop, as was the case for early phases of the CANARY instrument
\citep{canaryshort}, or the system can use \pol slope measurements
which are reconstructed from the residual slope measurements and
previous \ao system states (effectively the shape of the deformable
mirrors).  Given that all proposed \ao systems for the \elts operate
in closed-loop (or partial closed-loop in the case of MOSAIC
\citep{Hammer2014}), the latter technique must be used.

\subsection{Pseudo open loop slope computation}
In a closed-loop \ao system, the wavefront sensors measure the
residual wavefront slopes, i.e.\ the wavefront after correction by the
\dms.  Fortunately, when \dms are linear (or can be linearised), the
shape of the \dm can be estimated based on the \dm demands from the
previous frame (or frames, in the case of non-integer frame delay).
To compute the \pol slopes, the known \dm shape is used to reconstruct
the corresponding slope measurements that would be present if the \dm
surface was flat.  This operation
typically involves the multiplication of an interaction matrix with
the \dm demands.  These slope measurements are then added to the
residual slope measurements obtained from the wavefront sensors to
give the \pol slopes:
\begin{equation}
s^{\rm POL}_n = s^{\rm RES}_n + P\cdot a_{\rm n-1}
\end{equation}
where $s$ are the wavefront slopes (\pol and residual respectively),
$P$ is the interaction matrix (which can be measured in a conventional
way by poking the \dm), and $a$ are the actuator demands from the
previous frame (n-1).  In the case where the \ao loop delay is not equal to
one frame, a linear combination of actuator demands from neighbouring
frames should be used:
\begin{equation}
a = (1-d)a_{\rm n-D} + da_{\rm n-(D+1)}
\end{equation}
where $d$ is the fractional frame delay (between 0 and 1) and $D$ is
the total number of frames delay rounded down to the nearest integer.
The total delay is therefore given by $d+D$.  

Although for some \ao systems, $P$ can be sparse, for the \eso \elt it
is a dense matrix, of size equal to the total number of slope
measurements by the total number of actuators.  This is because the
\elt M4 mirror is treated as a modal \dm, with internal electronics
converting the applied modes (which we here call actuators for
cohesiveness) into actual actuator values (after possible
modifications in applied mode strengths to reduce stresses on the
mirror surface).  Therefore the
operation $P\cdot a$, which must be computed every frame, is a large
dense matrix vector multiplication, requiring significant memory
bandwidth (since the matrix has to be loaded from memory into the
computational units every \wfs frame, as it is too large to remain in memory cache).  This operation, along with wavefront
reconstruction, represents a bottleneck in the \ao real-time control
pipeline.  The \pol computation therefore represents a requirement for
a significant additional hardware cost in the \ao system design.  We
note that $P$ is the same size as the reconstruction matrix, with
required memory bandwidth scaling as the fourth power of telescope
diameter.

In \S2 we introduce a straightforward simplification that can be used
to greatly reduce the computational requirements of \pol computation,
whilst maintaining numerical exactness.  In \S3 we discuss the
implications that this method has, and additional requirements on the
\ao system supervisor.  We also consider several \ao system designs
and estimate the hardware savings that this method delivers.  We
conclude in \S4.

\section{Simplification of POL computation}

An \ao real-time controller updates the \dm demands at every \wfs time
step.  The \dm demands, $a$, required for a \dm are typically 
computed using
\begin{equation}
a_n = gR \cdot s^{\rm POL}_n + \left(1-g\right)a_{n-1}
\label{eq:orig}
\end{equation}
where $R$ is the reconstruction matrix (of size equal to the number of
actuators by the number of slope measurements) and
$g$ is the loop gain.  Expanding $s^{\rm POL}$ gives
\begin{align}
a_n &= R \cdot s^{\rm RES}_n &+ R\cdot P \cdot a_{n-1} &+
\left(1-g\right)a_{n-1}\\
    &= gR \cdot s^{\rm RES}_n &+ Q \cdot a_{n-1} &+
    \left(1-g\right)a_{n-1}
\label{eq:new}
\end{align}
where $Q$ is the precomputed matrix multiplication product of $R$ and $P$ multiplied by
$g$, a square matrix of size equal to the number of actuators squared.  It should be noted that in this latter equation, the \pol slopes
are never explicitly computed, and therefore not available to the
real-time control system.  We hereafter refer to this approach as {\it
  Implicit \pol}, while the conventional approach is referred to as
{\it Explicit \pol}.  This approach was first suggested by
\citet{2012SPIE.8447E..23W}, though did not include the gain term.  

We can therefore consider the differences in computational
requirements between these equations.  In each, two matrix-vector
multiplication operations are required.  In the original equation,
Eq.~\ref{eq:orig}, computation of $s^{\rm POL}$ requires
multiplication of the matrix $P$ (of size equal to the number of
slopes by the number of actuators) by a vector of size equal to the
number of actuators, followed by multiplication of $R$, (of size equal
to $P^T$) by s$^{\rm POL}$ (of size equal to the number of slopes).
The proposed version, Eq.~\ref{eq:new}, requires multiplication of $Q$
(a square matrix of size equal to the number of actuators), with the
actuator vector, and then a multiplication by $R$.  These matrices are
summarised in table~\ref{tab:sizes}
\begin{table}
  \begin{tabular}{llll}
    Matrix & Dimensions & Explicit POL & Implicit
    POL \\ \hline
    R & $n_{\rm act} \times n_{\rm slopes}$ & Yes & Yes\\
    P & $n_{\rm slopes} \times n_{\rm act}$ & Yes & No\\
    Q & $n_{\rm act} \times n_{\rm act}$ & No & Yes\\

  \end{tabular}
  \caption{A table summarising matrix dimensions for the implicit and
    explicit POL cases, with $n_{\rm act}$ giving the total number of
    actuators and $n_{\rm slopes}$ giving the total number of slopes.
  It should be noted that the number of slopes can be significantly
  larger than the number of actuators.}
  \label{tab:sizes}
\end{table}

\subsection{Real-time control system designs}
In order to compute the memory bandwidth requirement reduction that
can be made by implicit \pol computation, and hence the reduction in
hardware requirements, we must consider designs for some proposed \ao
real-time control systems.  We concentrate on \elt systems, including
\scao, \ltao, \mcao and \moao, mapping these designs to currently
available computational hardware.  We note that the \tmt architecture
\citep{10.1117/12.2314226} uses an explicit \pol computation.  A flexible
real-time controller, such as \darccite, will be able to use both
explicit and implicit \pol computation, given sufficient underlying
computational hardware.

\subsubsection{ELT SCAO}
\begin{figure}
  \includegraphics[width=\linewidth]{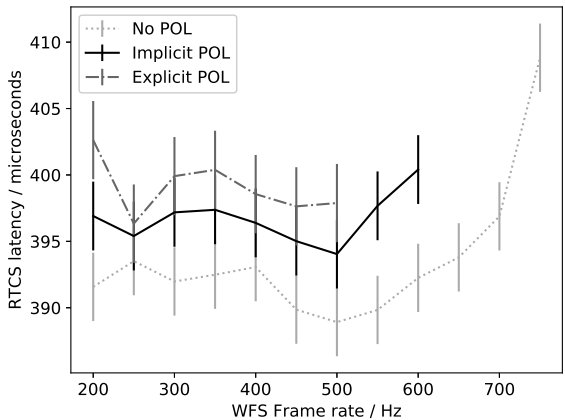}
  \caption{A figure showing \ao system latency as a function of \wfs
    frame rate for explicit and implicit POL computation.  For
    comparison, the case without POL is also shown, though this would
    result in lower \ao performance, since minimum variance
    reconstruction could not be used.  The explicit POL case cannot
    operate at frame rates above 500~Hz, while using implicit POL can
    extend the maximum frame rate to above 600~Hz.}
  \label{fig:scaoResults}
\end{figure}

The \elt \scao real-time control system will be comprised of a single
computational node, which receives \wfs pixels, and computes the
corresponding \dm demands.  When \pol is not performed, it has been
shown that a single Xeon Phi can process \elt \scao at about 1.2~kHz
\citep{jenkins2018}, in excess of the 1~kHz instrumental requirement.  
Table~\ref{tab:mvmsize} gives the memory bandwidth requirements for
both explicit and implicit \pol computation, and
Fig.~\ref{fig:scaoResults} shows measured \scao latency as a function
of \wfs frame rate for the explicit and implicit \pol cases, and for
the case without \pol calculation.  We find that the maximum \wfs
frame rate than can be processed using the explicit \pol calculation is
500~Hz, while when using implicit \pol calculation, this increases to
600~Hz, due to the reduced memory bandwidth requirement.
The latency is also lower.  Of course, the
case without \pol computation can reach highest frame rates (we show
up to 750~Hz for the camera model that we use here), since
memory bandwidths are significantly reduced.  However, in this case,
the \ao control algorithm is sub-optimal, since minimum variance
reconstruction cannot be used.  These measurements were taken using
\darccite, on an Intel Xeon Phi 7250 processor.  The configuration of
\darc is described by \citet{jenkins2019}.  In particular, we use a
camera simulator to provide pixels in the same format as the \eso \elt
wavefront sensors. 

\begin{figure}
  \includegraphics[width=\linewidth]{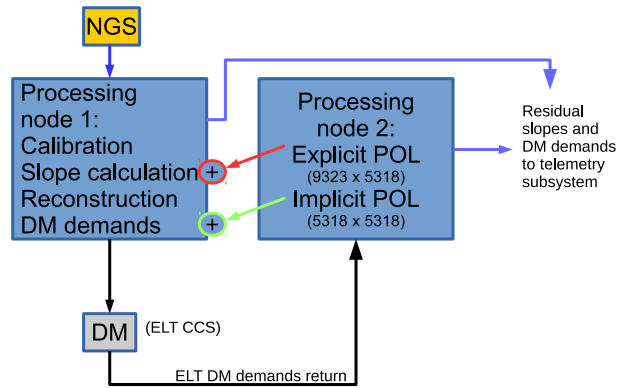}
  \caption{A possible ELT SCAO real-time control system architecture
    based on two CPU nodes.  Explicit POL computation data flow is denoted by
    red arrows, while implicit POL computation data flow is denoted by
    green arrows.  The explicit POL computation result is required earlier in
  the pipeline, and therefore has more stringent latency
  requirements.  Pixel and telemetry information flow is represented
  by blue arrows while DM demands are represented by black arrows.}
  \label{fig:scao}
\end{figure}

For higher frame rates, the Xeon Phi used here is unable to process
\wfs information fast enough, and therefore dropped frames and
eventual real-time control system failure result due to the increased
latency.  Although the memory bandwidth requirements are theoretically
achievable using a single Xeon Phi processor (480~GBs$^{-1}$), or a
quad-CPU Intel Scalable Processor design (512~GBs$^{-1}$), in
practice, it is not possible to compute the \pol computation within a
single node while achieving higher frame rates or reduced latency, due
to the other operations necessary (pixel reception, calibration, etc).
In particular, the Xeon Phi has poor single thread performance, so
aspects of the real-time control system which cannot be parallelised,
such as pixel acquisition, can significantly impact performance.
Therefore, a second computational node could be used to receive the
applied \dm demands (as sent from the \elt \ccs), and compute the
component of the implicit or explicit \pol slopes.  This node would
then send either slope adjustments (explicit \pol) or actuator
adjustments (implicit \pol) to the first computational node, as shown
in Fig.~\ref{fig:scao}.

In this case, the use of implicit \pol does not reduce the required
hardware (2 computational nodes are still required).  However, \ao
system latency and jitter (variation in latency) can be improved since
the implicit matrix-vector operation is smaller, and the result
combined later, i.e.\ added to actuators after residual wavefront
reconstruction, rather than being added to residual slopes before
wavefront reconstruction.  We note that a hardware accelerator, for
example a \gpu could be used in place of the second node, though here
we consider only CPU-based designs.

In the case that there is a variation in arrival times of the \dm
demands sent from the \elt \ccs (e.g. due to the network, or
non-deterministic algorithms), the implicit \pol method has a larger
computational window available to mitigate the effect of this jitter.

\subsubsection{ELT LTAO}
\begin{figure}
  \includegraphics[width=\linewidth]{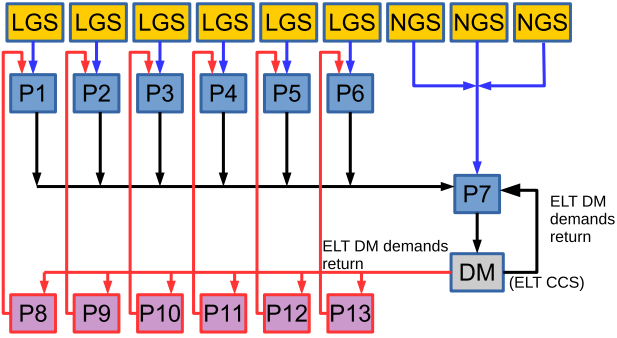}
  \caption{A possible ELT LTAO real-time control system architecture
    based on seven CPU nodes (or 13, in the case of explicit POL
    computation).  Explicit POL computation data flow and extra
    components are shown in red (requiring six extra nodes, P8--P13).
    P represents a processing node (numbered from 1 to 13), with nodes
    1--7 performing image calibration, slope calculation and partial
    wavefront reconstruction, node P7 additionally performing DM
    vector summation, and nodes 7--13 computing POL information (all
    in node 7 in the case of implicit POL).  Blue arrows represent
    pixel data, black arrows show DM data.  }
  \label{fig:ltao}
\end{figure}

The proposed \elt \ltao system uses 6 \lgss and between one to three
low order \ngss to control the telescope deformable mirror via the
\ccs.  Fig.~\ref{fig:ltao} shows a schematic design for the real-time
control system, with each \wfs being sent to a single processing node.
These nodes each perform image calibration, residual slope calculation
and partial wavefront reconstruction.  Finally, these nodes compute partial \dm vectors which are then collected by
the \ngs node (i.e.\ a ``gather'' node) and summed together to give
the final \dm demands, before being sent to the \dm.

In the case of implicit \pol reconstruction, the \pol computation is
performed on this \ngs node, and therefore only this node needs to
receive the applied \dm demands from the \ccs.  This \pol calculation
involves a square matrix-vector multiplication with dimensions equal to
the number of actuators.

When using implicit \pol calculation, the operations performed on
nodes P1--P6 are essentially very similar in form to a \scao
calculation without any \pol, i.e.\ a residual slope computation,
and reconstruction using these.  Fig.~\ref{fig:scaoResults} shows that
using Xeon Phi nodes, frame rates in excess of 750~Hz could be achieved here.

When using explicit \pol calculation, on the \wfs nodes, image
calibration and residual slope computation are performed.  The
 \pol residuals are then added, followed by
partial wavefront reconstruction, before computing the partial \dm
demands which are then collected and summed by the \ngs node.  These
\dm demands are sent to the \elt \ccs.  The \elt \ccs will then return
the actually applied demands.

If these applied demands are sent to the \wfs nodes, P1--P6, the \pol
calculation for each \wfs is then performed, so that the \pol slope
residuals can be added to the residual slope measurements.  We note
that the size of these operations on each \wfs node is essentially
identical to that of a \scao system, and therefore, as shown in
Fig.~\ref{fig:scaoResults}, a maximum frame rate of about 500~Hz will
be achievable using Xeon Phi nodes.

In the likely case that higher frame rates will be required, the
applied demands can be sent from the \ccs to additional nodes,
P8--P13.  On these nodes, \dm demands are then
multiplied by an interaction matrix to give the explicit \pol slope
residuals, to be send to the \wfs nodes for addition during the next frame.

The additional communications required for the explicit \pol
computation will reduce reliability.  If a UDP protocol is used
(including multicast), packets can go missing, while using a TCP
protocol will increase jitter.  In some situations, where delays mean
that \pol slope residuals are computed very close to the time at which
they are required, timing issues may result, with different \wfss
either being delayed, or using measurements from a previous frame.
With the implicit \pol approach (which is significantly simpler), this
will not happen, since all \pol computation is performed within a
single node.  Simplicity within a real-time control system design is
advantageous.  

We note that in the explicit \pol case, rather than using additional
nodes to achieve higher frame rates, it would also be possible to use
different computational hardware.  For example a quad-CPU Intel Xeon
Gold or Platinum system would achieve slightly higher memory bandwidth
than a Xeon Phi, with improved single-core performance. Unfortunately,
the cost of a single such node would be at least double that of two
Xeon Phi nodes, based on current prices.  We also note that
we find that theoretical performance metrics do not reliably relate to
full real-time control system performance, when real camera packets
are included.

In the case of explicit \pol reconstruction, 6 additional nodes are
required to perform this calculation (shown in red) before sending the slope
adjustments to the corresponding \wfs processing node.  Each of these
additional nodes must also receive the applied \dm demands from the
\ccs.  This therefore represents a significant increase in complexity
when compared with the implicit \pol approach.

Telemetry information is not shown in Fig.~\ref{fig:ltao} for
clarity.  However, the necessary telemetry information (discussed
further in \S\ref{sect:telem}) includes
residual slopes from nodes P1--6, \pol slopes from nodes P1--6 when
using explicit \pol computation, and \ccs return demands from P7.

We note that while a single node based on currently available hardware
(e.g.\ a Xeon Phi, AMD EPYC or Intel Scalable processor solution) may
be able to process a single \wfs and its corresponding \pol
calculation (in the explicit case), such a solution would likely be
unstable, with increased jitter and the risk of dropped frames,
particularly at moderate frame rates (e.g.\ 500~Hz) and above.  It
would also be more susceptible to problems arising from the timing of
arrival of \ccs information.  For this reason, Fig.~\ref{fig:ltao}
adds additional nodes for the \pol computation.

\subsection{ELT MCAO}
\begin{figure}
  \includegraphics[width=\linewidth]{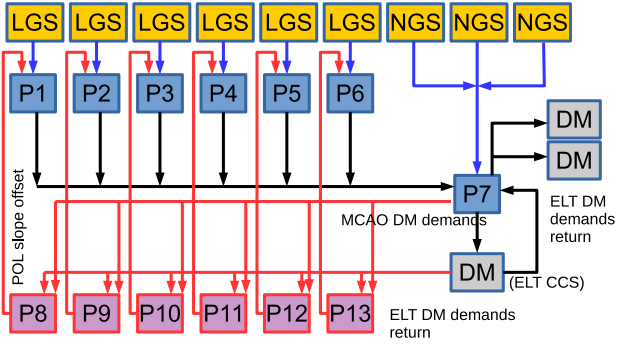}
  \caption{A possible ELT MCAO real-time control system architecture
    based on seven (or 13) CPU nodes.  Explicit POL computation data
    flow and extra components are shown in red (requiring six extra
    nodes, P8--P13), and DM demands are required to be sent from the
    CCS and from node P7 to the additional nodes.  P represents a
    processing node (numbered from 1 to 13).  For the two additional DMs
    it is assumed that the requested and applied DM demands are equal
    (as is usually the case).   Blue arrows represent
    pixel data, black arrows show DM data.  }
  \label{fig:mcao}
\end{figure}

From the point of view of the real-time control system, the \elt \mcao
case is similar to the \ltao case.
However, the differences are that the \mcao system will have an
increased number of actuators to compute due to the two additional
non-zero-conjugate \dms.  We assume here that the \dm demand return will not be returned
from these \dms (as was the case for the \ccs), but will be computed
and known by node P7, as shown in Fig.~\ref{fig:mcao}.  Therefore, in the case
of implicit \pol, these demands are not sent to other nodes, and are
used for implicit \pol computation on node P7.  However,
in the case of explicit \pol, these demands will be sent from P7 to
nodes P8--13, so that the explicit \pol computations can be
performed.  This represents an additional increase in complexity, and
greater potential for instability, particularly in the case of
real-world situations where Ethernet packets can get lost.

\subsection{ELT MOAO}
\begin{figure}
  \includegraphics[width=\linewidth]{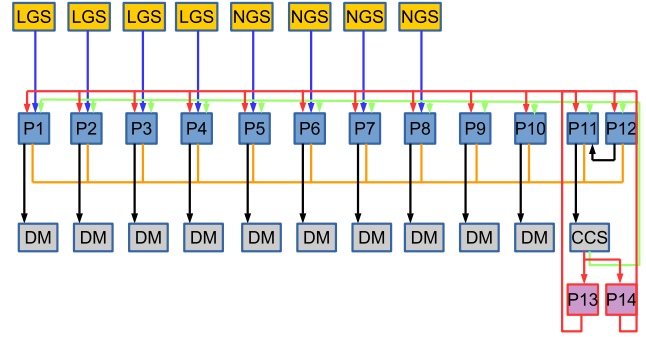}
  \includegraphics[width=\linewidth]{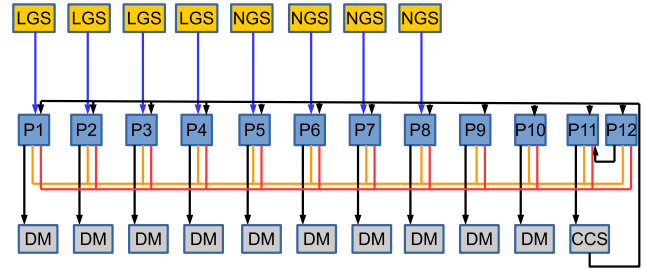}
  \caption{(a) A possible ELT MOAO real-time control system architecture
    based on twelve CPU nodes (or 14 nodes when explicit POL
    computation is required).  The extra components and data flow
    required for explicit POL computation are shown in red.  The
    orange lines represent the multicast of slope measurements between
    nodes.  Green lines represent distribution of CCS demand output
    when implicit POL is used.  Blue lines represent pixel flow and
    black lines represent DM demand flow.
    (b) An alternative ELT MOAO architecture, using 12 CPU nodes in
    both cases.  Here, in the explicit case, partial POL
    reconstruction is performed on each node (using the CCS demands).
    The resulting slopes are then broadcast to all other nodes (red
    lines).  Computation of POL slopes is shared equally between
    nodes, as otherwise, the bandwidth requirements for P11 and P12
    are likely to be too great.
  }
  \label{fig:moao}
\end{figure}

The proposed \elt \moao system uses 4 \lgs and 4 high order \ngs.  In
addition to the \ccs actuators, ten open-loop \dms are also used (with
a lower spatial order than the \ccs mirror, having $48\times48$
actuators.  Fig.~\ref{fig:moao} shows possible schematic designs for
the real-time control system.  In each case, images from each \wfs are
sent to a single processing node, which computes the corresponding
wavefront slopes.  These slopes are then distributed (multicast) to
all other nodes, so that all nodes now have all slope measurements.
Wavefront reconstruction is performed (in a pipelined fashion, as soon
as the corresponding slopes become available) on each node which has
sole responsibility for commanding a single \dm.  The only caveat here
is that control of the \ccs is shared by two nodes due to the high
memory bandwidth requirement, with node P12 passing its computation
back to P11 so that the final \ccs demands can be sent.  We propose
the use of two nodes for this task, since a minimum of 327~GBs$^{-1}$
is required (to be able to compute the reconstruction at loop rate,
with higher bandwidth being necessary for reduced latency), which
results in too low spare overhead for a single node, risking system
instability and increased latency.

For implicit \pol computation, the applied \ccs demands are sent to
all processing nodes where the POL calculation is performed.

For explicit \pol computation, there are two possibilities.  In the
first case (Fig.~\ref{fig:moao}(a)), two additional nodes are used for
the explicit \pol calculation (with the same memory bandwidth
requirements as nodes P11 and P12), with these slopes then being
broadcast to all other nodes, which add to the residual slopes to give
the \pol slopes.  This results in extra cost and complexity compared
with the implicit calculation where the \ccs output is sent directly
to the processing nodes.

In the second case for explicit \pol (Fig.~\ref{fig:moao}(b)), the
applied \ccs demands are sent to all nodes, which then equally share
the computation of \pol slopes.  These \pol slopes are then
distributed to all other nodes.  In this case, the number of residual
slopes and \pol slope offsets computed on each node differs and so a
separate broadcast of each is required, reducing reliability.

We also note that an 8 node design would also be possible here.  In
this case, since each \moao \dm is independent, the 22666 actuator
values would need to be spread between these nodes (approximately 2834
actuators per node), and then broadcast and collected.  These nodes
would then need to send \dm demands to one or two \dms each.  Since
\dm demands cannot be broadcast until all computation is completed
(i.e.\ all slopes have been reconstructed), latency is increased
compared with our first approach with one \dm per node, where slopes
can be broadcasted as they become available, in a pipelined fashion.

\subsection{Summary of POL operations}

Table~\ref{tab:mvmsize} summarises the
relative size of these operations for different proposed \ao systems.

\begin{table*}
  \begin{tabular}{llllll}
AO system & Number of & Number of  & Frame rate &
Bandwidth for  & Bandwidth for \\
&slopes &actuators&&Explicit POL & Implicit POL\\ \hline
ELT SCAO & 9232 & 5318 & 1~kHz & 393~GBs$^{-1}$ & 310~GBs$^{-1}$ \\
ELT LTAO & 55392 & 5326 & 500~Hz & 1180~GBs$^{-1}$ & 647~GBs$^{-1}$ \\
ELT MCAO & 55392 & 6326 & 500~Hz & 1402~GBs$^{-1}$ & 781~GBs$^{-1}$ \\
ELT MOAO & 61504 & 22666 & 250~Hz & 1722~GBs$^{-1}$ & 1515~GBs$^{-1}$\\
TMT NFIRAOS & 35808 & 7675 & 800~Hz & 1859~GBs$^{-1}$ & 1068~GBs$^{-1}$\\ \hline
  \end{tabular}
  \caption{A table comparing real-time controller memory bandwidth
    requirements for the implicit and explicit \pol computations.
    These numbers also include the wavefront reconstruction.  LTAO and
  MCAO slope count does not include the NGSs, but as these are low
  order, the results are not affected significantly.}
  \label{tab:mvmsize}
\end{table*}

It can be seen that in the \ltao and \mcao cases, the memory bandwidth
requirement is reduced by nearly 50\% when using the implicit \pol
method.  

For the \moao case, we have assumed 4 \lgs and 4 \ngs each with
$74\times74$ sub-apertures giving a total of $N_\mathrm{s}=61504$ slopes, which is the current MOSAIC baseline
\citep{mosaicbaseline}.  We also assume that there are 10 \moao \dms each with $A_\mathrm{M}=1634$ actuators ($48\times48$), in
addition to the \elt \ccs (with $A_\mathrm{CCS}=5326$ actuators).

For the explicit \pol \moao case, the memory bandwidth required is
therefore
\begin{equation}
\left( \left( A_\mathrm{CCS} + 10 A_\mathrm{M} \right) \times
N_\mathrm{s} + A_\mathrm{CCS} \times N_\mathrm{s} \right) \times 4 f
\end{equation}
where $f$ is the frame rate, and the factor of 4 is the number of
bytes in a 32-bit floating point integer.  The left part of this
equation represents the residual slope reconstruction, while the right
part ($A_\mathrm{CCS} \times N_\mathrm{s}$) represents the explicit \pol slope computation.  The explicit
\pol slope computation can be performed either on two separate nodes
(Fig.~\ref{fig:moao}(a), or spread over the existing nodes
(Fig.~\ref{fig:moao}(b).  

For the implicit \pol \moao case, the memory bandwidth is
\begin{equation}
\left( \left( A_\mathrm{CCS} + 10 A_\mathrm{M} \right) \times
N_\mathrm{s} + A_\mathrm{CCS}^2 + 10 A_\mathrm{CCS}
\times A_\mathrm{M} \right) \times 4f
\end{equation}
where the left part of this equation represents the residual slope
reconstruction, and the right part ($A_\mathrm{CCS}^2 + 10 A_\mathrm{CCS}
\times A_\mathrm{M}$) is the implicit \pol computation.
For the implicit case, the memory bandwidth required for the \pol
computation is approximately a third of that required in the explicit
case.

\subsubsection{Considerations of sparsity in the interaction matrix}
Even in the case where $P$ can be sparse, the method that we propose
here is still relevant, particularly for \mcao and \ltao cases, as it
reduces communication complexity, since it removes the need to
send the \pol slopes back to the \wfs nodes, thus reducing complexity
and jitter, and increasing reliability. We also note that with a sparse $P$
matrix, the post processing of closed loop residuals as a batch
process (described in \S~\ref{sect:batch}) to
give \pol slopes for tomography benefits from the sparsity
(i.e. becomes a cheaper operation), as does the Q=R*P matrix
multiplication product,
which then becomes achievable in shorter timescales or with reduced
hardware.

\section{Implications}
Our proposed implicit \pol method means that \pol slopes are not
explicitly computed by the real-time control system.  However, \pol
slopes are a necessity for parts of the \ao system (such as
tomography), and therefore we discuss these requirements, and
solutions here.

\subsection{Computation of Q}
The implicit \pol technique requires the computation of $Q$ in
Equation~\ref{eq:new}.  This operation is a matrix-matrix
multiplication, with the matrix sizes equal to the number of actuators
by the number of slopes.  For the \elt \scao case, we have $Q=R\cdot
P$ with $R$ having dimensions $5318\times 9232$ and $P$ having
dimensions $9232\times 5318$.  This can be performed in under a second
on a standard Xeon server (dual E5-2630-v3 CPUs dating from 2015), and
therefore would not require any additional hardware beyond what would
be expected for an \ao supervisor hardware system (used for
calibration and soft real-time tasks).

For the \ltao case, these matrices increase in size to $55392\times 5326$,
and the computation of $Q$ takes less than one minute on this server.
On a modern \gpu, this would be far faster, and nearly trivial to
implement (e.g.\ a call to a BLAS library).  It is also a highly
parallelisable problem and so could be spread across several CPU nodes
using the Message Passing Interface (MPI) as required.  However, an
update rate of once per minute is likely to be sufficient, so this is
not a difficult operation.  It is also not an operation within the
real-time pipeline, so can be performed on hardware using a standard
(non-real-time) operating system, and considerations of jitter are not
necessary.  The \mcao case is similar.

For the \ltao and \mcao cases, if a faster update rate is required,
additional computational hardware would be necessary in the implicit
\pol case.  However, this would be less than the additional nodes
required to implement explicit \pol (which would also introduce extra
complexity and risk, including the necessity to pass \dm demands back
to the \wfs nodes).

\subsection{Atmospheric statistic computation}
\label{sect:batch}
Computation of atmospheric statistics is necessary for building the
minimum variance control matrix, which requires a knowledge of the
C$_n^2$ profile, and estimates of Fried's parameter and the outer
scale.  This requires open-loop slope measurements \citep{polc}.
Fortunately, this information is not required in real-time on a frame
by frame basis.  Therefore, by recording contiguous frames of residual
slopes and \dm demands (a standard operation for facility \ao
systems), the \pol slopes can be reconstructed after the event.  The
key advantage here is that \pol slopes can be computed as a batch,
i.e.\ several thousand frames of \pol slopes can be computed
simultaneously:
\begin{equation}
S^{\rm POL} = S^{\rm RES} + P \cdot A
\end{equation}
where $S$ is a matrix comprised of many frames of slopes (dimensions
equal to number of slopes by number of frames) and $A$ is a
matrix comprised of many frames of \dm demands (dimensions equal to
number of actuators by number of frames).

\begin{figure}
  \includegraphics[width=\linewidth]{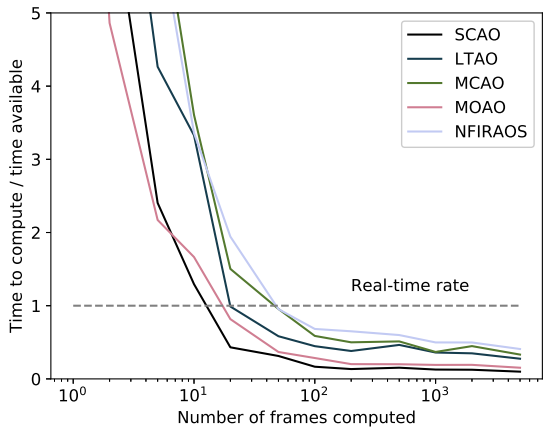}
\caption{A figure showing the ratio of time to compute a batch of POL
  slopes to the time within which these slopes are produced, as a
  function of number of frames.  When this ratio falls below unity,
  production of POL slopes is able to keep up with the real-time
  control system.}
\label{fig:batch}
\end{figure}

This calculation involves a matrix-matrix multiplication, rather than
a matrix-vector multiplication, and therefore is a compute-bound
rather than a memory-bound operation.  We have bench-marked this
operation using a conventional server computer, with dual Xeon
E5-2630-v3 CPUs (circa 2015), and find that a single server is able to
keep up with the real-time data rates produced by the real-time
control system when blocks of at least approximately 100 frames of
slopes are processed simultaneously, as shown in Fig.~\ref{fig:batch}.
This corresponds to a delay of less than a second between implicit and
explicit \pol slopes, and this does not pose any problems for the \ao
control system.  Table~\ref{tab:after} summarises the benchmark
timings for the different cases given in table~\ref{tab:mvmsize}.  In
all cases, it is possible to process one second of data in less than
one second.  On more modern hardware, processing time would be
significantly reduced due to increased width vector processing units
(e.g.\ AVX512).  Fig.~\ref{fig:batch} shows the transition between
memory-bound operation (for small batch sizes) to compute-bound
operation.  This figure shows the time taken to compute a batch of a
given number of frames, and therefore the latency of this computation
(from first frame to end of computation) will equal the time taken for
the real-time control system to deliver this batch (i.e.\ frame period
multiplied by batch size) plus the batch computation time.  

\begin{table*}
  \begin{tabular}{lllll}
AO system & Rate for 1000 frames & Rate for 5000 frames & Time to
process 1s data &  MFLOP/frame \\ \hline

ELT SCAO & 7.6~kHz & 8.9~kHz & 0.13s & 98\\
ELT LTAO & 1.4~kHz & 1.7~kHz&  0.45s &590\\
ELT MCAO & 1.1~kHz & 1.5~kHz & 0.67s&701\\
ELT MOAO & 1.2~kHz & 1.6~kHz & 0.24s&655\\
TMT NFIRAOS & 1.5~kHz & 1.9~kHz & 0.52s&550\\ \hline
  \end{tabular}
  \caption{A summary of computation time for explicit \pol slopes
    derived from residual slopes and DM demands.  These are computed
    using the SGEMM function available in the Intel Math Kernel
    Library.}
  \label{tab:after}
\end{table*}

We do not anticipate that additional hardware would be required for
this task, as it could be performed by supervisory sub-system hardware
(i.e.\ a soft-real-time system used for calibration and system
supervision).  Since this is not a hard real-time task, jitter in this
operation is not critical.

\subsection{Telemetry requirements}
\label{sect:telem}
Using our implicit \pol technique, to compute explicit \pol slopes
after the event, the \ao system must record continuous frames of
residual slopes and \dm demands (actually applied, i.e.\ as returned
from the \ccs).

If an explicit \pol technique is used, the \ao system will usually
record both residual and \pol slopes (residual slopes being useful for
telemetry displays and for computation of telescope offloading).  In
addition to this, \dm demands would also usually be recorded, though
it would be possible to compute these from the residual and \pol
slopes.  In any case, the explicit \pol technique requires capture of
more telemetry information than the implicit case.

\subsection{Simplification of control system network}
Another significant benefit to the implicit \pol technique is a
simplification of the real-time control system network requirements.
As can be seen in Figs.~\ref{fig:scao},\ref{fig:ltao},\ref{fig:mcao} and
\ref{fig:moao}, fewer data paths between the real-time processing
nodes are required.  This increases system robustness, since there are
fewer packets that could get lost (and we assume that UDP packets will
be used due to the stringent latency requirements ruling out any TCP
retransmissions).

\subsection{Slope linearisation}
Shack-Hartmann wavefront sensors are non-linear due to pixelisation
effects.  Although this effect is small, and usually ignored, under
certain conditions, \ao performance improvements can be achieved by
using a look-up table or polynomial fit to linearise slope
measurements \citep{basden8}.  Solar \ao wavefront sensors can be
particularly affected by this non-linearity.

Once the residual slopes have been linearised it is not necessary to
take any further linearisation steps with either explicit or implicit
\pol schemes.

\section{Conclusions}
We have presented a straightforward and mathematically equivalent
technique for the implicit computation of \pol slopes for minimum
variance \ao systems, and considered the implications for common \ao
operation modes.  This technique simplifies real-time control system
design (hence reducing cost), increases robustness and reliability of
the \ao system, and leads to a reduction in required telemetry
information.  It leads to a simplified real-time control pipeline,
which leads to lower latency and jitter.  However when using the
implicit scheme, if explicit slopes are required (for example, for
atmospheric parameter estimation) an additional non-real-time
computation is required, though we note that this still results in a
lower server requirement (and hence cost) than the traditional
explicit \pol approach, and can be performed using a relatively
low-end server, or shared with an existing \ao supervisor server,
e.g.\ that used for control matrix computation.  Therefore,
significant hardware and complexity savings can be achieved using
implicit \pol computation.

\section*{Acknowledgements}
This work is funded by the UK Science and Technology Facilities
Council grant ST/K003569/1 and consolidated grants ST/L00075X/1 and
ST/P000541/1.  This work is also sponsored through a grant from project
671662, a.k.a. Green Flash, funded by European Commission under
program H2020-EU.1.2.2 coordinated in H2020-FETHPC-2014.

\bibliographystyle{mn2e}

\bibliography{mybib}
\bsp

\end{document}